\documentclass[12pt]{article}
\usepackage{amssymb}
\usepackage{amsmath}

\def\beq{\begin{equation}}\def\eeq{\end{equation}}
\def\bea{\begin{eqnarray}}\def\eea{\end{eqnarray}}

\begin{document}
 
\title{Quantum mechanics and gravity as preclusion principles of four dimensional geometries}
 
\author{Roman Sverdlov
\\Physics Department, University of Michigan,
\\450 Church Street, Ann Arbor, MI 48109-1040, USA}
\date{October 11, 2008}
\maketitle
 
\begin{abstract}

\noindent The goal of this paper is to employ a "preclusion principle" originally suggested by Rafael Sorkin in order to come up with a relativistically covariant model of quantum mechanics and gravity. Space-time is viewed as geometry as opposed to dynamics, and "unwanted" histories in that geometry are precluded. 

 \end{abstract}

\noindent{\bf 1. Introduction}

Rafael Sorkin in his work on interpretation of quantum mechanics (see [1]) has proposed an idea of preclusion, where global correlations of events that do not match a probability prescribed by quantum mechanics are being systematically "precluded". A toy example of "preclusion" would be replacing a statement "a dice has probability 70 percent of falling on side A and 30 percent of falling on side B" with a list of the "precluded" behaviors: we can say that if we throw a dice 100 times then whenever it falls less than $m$ times on side A or more than $n$ times on side B then that particular pattern is precluded, where $n < 70 < m$. According to this model, all of the non-precluded behaviors of a dice are equally likely, while all the precluded ones are equally forbidden, regardless of their actual probabilities. We can envision this model as a collection of parallel universes where we have exactly one universe for every non-precluded behavior.

While I don't know for sure, to the best of my knowledge, Sorkin was using the preclusion principle dynamically: he imagined universe to be dynamically growing, and he obtained more and more precluded states as that growth continues. On the other hand, in this work I would like to apply his principle to the spacetime manifold the way it is imagined in a classical gravity: a static picture of four dimensional object where time is solely a part of geometry and there is no room for any kind of evolution, either classical or quantum mechanical. I will do that by claiming that "wrong" behavior observed by any observer, regardless of their reference frame, is equally a ground for preclusion. Thus, the natural selection that selects the "right" quantum correlations would be frame-independent.  

According to my model, there is no such thing as an observer. Fields are well defined throughout space-time and they are all observed. However, they are adjusted in such a clever way that if some observer will "decide" not to observe fields in a particular region, statistically he would observe the global correlations of fields on the boundary of that region to be the ones expected by quantum mechanics. There is no dynamical law that would make that happen. Instead, we simply "preclude" all the behaviors of the fields where this is not approximately the case. 

While the rigorous work is still ahead, the qualitative argument is made that the "preclusions" can be made self-consistent if we impose a lower bound on an entropy density of the universe, so that decoherence will always occur on a certain large scale.  This might be considered a side benefit since this implies third law of thermodynamics in a natural way. 

This should be contrasted with Bohm's Pilot Wave model. On the one hand, in that model there is a precise evolution equation of beables from which the desired quantum correlation follows as a result. At the same time, however, that guidence equation is written in specific frame, and thus violates a covariance principle. On the other hand, in our work we made the laws less "rigid" by allowing more than one behavior -- in fact we allow all non-precluded behaviors. This gives more flexibility to allow our version of dynamics to simultaneously be true in all frames. The price to pay, which is lack of determinism, can be dealt with by imaginning uncountably many parallel universes in which every single non-precluded behavior is realized. 

Finally, I apply the preclusion principle to gravitation. I will illustrate how I can use that principle to develop two conflicting and competing models of gravity: A view originally expressed by Dyson (see [2]) according to which as long as graviton has never been observed, gravity can be viewed as solely classical phenomenon (section 5) and, in parallel, a model where gravity IS a quantum phenomenon, and in particular is quantized by means of path integrals on a causal set (section 6). The reason I pursue two opposite views is that I believe in intellectual pluralism. As long as there is at least one person on each side of a table, it means that you can't conclusively prove the validity of one view over the other. This means that the more views a theory accomodates, the "safer" theory it is.

The way I support Dyson's view is as follows: since the philosophy of my work is that specific physics is NOT dynamically generated but rather is selected out through preclusion of all other physics, I don't have to invent any microscopic dynamics of gravity that would "generate" Einstein's equation on a macroscopical level. Instead, I simply "preclude" any macroscopic behavior where Einstein's equation doesn't happen to hold; the way I make sure I only preclude macroscopic violations and not microscopic ones is to postulate an "approximate" Einstein's equation where the coefficients of $1$ and $1/2$ are only approximately equal to these values with a well defined tolerance of approximation.   

On the other hand, the way I allow the possibility of quantization of gravitational field is as follows: According to this work, all fields, gravity and otherwise, are well defined throughout space time. A quantization of any particular field does NOT imply that it fluctuates. Rather, it is a statement that if an observer choses to ignore its behavior inside a certain region, the correlations of a given field on the boundary of that region should match a quantum mechanical prediction. Since the pattern is pre-designed this way and is observer-independent, none of the "quantized" fields are uncertain. This means that we are free to include gravity in our list of "quantized" fields while still having a well defined geometry. 

The only way in which our version of Dyson's model (section 5) is different from our version of quantum gravity (section 6) is that in the former case the constraint involves an approximate validity of classical Einstein equation while in the latter case the constraint involves a correlation with path integrals of IMAGINED fluctuations over various regions. Neither of these constraints imply that the original gravitational field is uncertain. Thus, while I pursue both quantum-gravity and quantum-less-gravity views, unlike what one might expect, in both cases I have well defined non-fluctuation geometries some of which are being precluded.   

In section 2 of this paper I will restrict myself to toy model of absolute time. While doing so, I will introduce my own version of Sorkin's preclusion principle specifically designed in a way that would be suitable for the introduction of relativity in the rest of the paper. Then in section 3 I will introduce general relativistic covariance, but I will ignore the dynamics of gravitational field, whether it be Einstein's equation or the action for gravity. I will focus on behavior of other fields on a fixed curvature background. Then in section 4 I will show how to extend section 3 to fermionic cases. One key ingredient is a summary of paper [9] where I defined Grassmann variables and ferionic fields as literally existing outside path integration; this will take up a first half of section 4. The other half of section 4 is devoted to "localizing" fermions into world path, which will follow Bohm's tradition of using position beables for fermions and field beables for bosons. Finally, sections 5 and 6 will be devoted to gravity. Section 5 will use the preclusion principle to substantiate Dyson's view of gravity according to which gravity only exists classically. Section 6, on the other hand, will illustrate how the preclusion principle can provide a solid ground for causal set approach to quantizing gravity to work. Finally, section 7 will summarize my views of what was done so far and my ideas of possible directions for future research.     

\bigskip
\noindent{\bf 2. Toy model: non-relativistic quantum mechanics in absolute time}

As we have said, we would like to follow the idea expressed by Rafael Sorkin in his work on quantum measure that deals with forbidding the low probability histories. Essentially, that means that we forbid all the global histories in which global correlation of events does not indicate a desired probability. Thus, if we have a dice whose probability of landing on one side is 70 percent, we replace that statement with a statement that we forbid all the global histories of behavior of the dice where the ratio between its two outcomes differ from 70 percent by more than some allowed margin. Since we have abandoned the notion of probability, all the patterns that are within margin are equally allowed, despite the fact that their probabilities differ. Thus, it is not a "true" probabilistic process; however, it is globally taylored in such a way that an ignorant observer would still come up with a conclusion that it is one. We can further hypothesize an existence of parallel universes where every single "allowed" history is realized. This hypothesis restores determinism. 

An advantage of the theory is this: since we no longer believe in probability but rather in ruling out of the "unwelcome" patterns, the fact that the set of "unwelcome" patterns happen to coincide with the one we would expect from classical version of probability theory is mere coincident. Thus, we can replace the criteria of exclusion of histories by anything else we like that doesn't have to even resemble probability at all. In particular, we can replace a "classical" probabilistic correlation with quantum mechanical one without a burden of calling it probability or explaining how it works. Thus, the "exclusion" theory has two separate benefits. Apart from the fact that it removes the notion of probability with a well determined list of parallel universes, it also allows complex probabilities to make just as much sense as the real ones.

Now lets move on to trying to impliment it to quantum mechanics. Since I consider a situation where many particles are living in absolute time, we can consider a configuration space, so that different coordinates of a point in configuration space correspond to coordinates of different particles that co-exist "simultaneously" in time, where due to absolute time the notion of "simultaneously" is well defined. The potential corresponding to the interaction of particles with each other in regular space will correspond to the fixed potential imposed from outside on the phase space. In the phase space picture, we have a single particle interacting with that potential living in multidimensional space. 

According to my model, there are many parallel universes. In each of the parallel universes a particle is always localized, and moves along well defined trajectory, so that each of the "allowed" trajectories is realized in one of the parallel universes. Now, we have to come up with a criteria of rulling out forbidden trajectories. In order to avoid mathematical difficulties of defining the probabilities of patterns for infinite time interval, we will make sure that time interval is finite, say it goes from $t=-T$ to $t=T$. Suppose we have a trajectory of a particle, $\alpha(t)$ that is to be tested for whether or not it is allowed. We are going to test it by taking every possible time interval $(a, b)$. In each case, we will imagine some other curve, $\beta$, co-existing with a curve $\alpha$, where the curve $\alpha$ is the beable while the curve $\beta$ is quantum mechanical. I will then use the usual rules of quantum mechanics to compute probability amplitude of $\beta (b) = \alpha (b)$ given an initial condition of $\beta (a) = \alpha (a)$. This would be 

\bea {\cal A} (\alpha; a,b) = \frac{\int_{\alpha(a)= \beta (a); \alpha (b) = \beta (b)} [d \beta] e^{iS(\beta(a,b))}}{\vert \int_{\alpha(a) = \beta (a)} [d\beta]{ e^{iS(\beta (a,b))}} \vert  } \eea

and the probability density is 

\bea p (\alpha; a, b) = \vert {\cal A} (\alpha; a, b) \vert ^2 \eea

We will now would like to impose a constraint of the form 

\bea p_0 - \epsilon < \mu \{ (a,b) \vert p_0 - \delta < p (a,b) < p_0 + \delta \} < p_0 + \epsilon \eea

 We might and might not want to make that criteria stronger by separately imposing the probability cutoffs for each time interval $\delta t = b-a$

\bea \forall \delta t > \Delta (p - \epsilon < \mu \{ (a,b) \vert b-a= \delta t \; and \; p_0 - \delta < p (a,b) < p_0 + \delta \} < p_0 + \epsilon ) \eea

Here, $\Delta$ is a function of the lowest attainable temperature defined in such a way that in the limit of lowest possible temperature going to $0$, $\Delta$ approaches infinity. The fact that I still have restrictions for time intervals greater than $\Delta$ implies that measurements still occur on these scales, even if we don't have a complex system. At the same time, the fact that the postulate is applicable not only to $\Delta$ but also to, for example, $2\Delta$ implies that fields should, somehow, be adjusted in such a way that two measurements with an interval $\Delta$ and one measurement with an interval $2 \Delta$ produce the same result. For example, we can have a double slit experiment where the distance between emission of a particle and either of the two slits is $\Delta$ and the distance between these slits and the screen is also $\Delta$. In this case, the constraint applied to $\Delta$ will tell us that to compute probabilities of a particle reaching a screen we have to simply add probabilities, rather than probability amplitudes, of particle going through either of the two slits; on the other hand, if we apply that constraint to $2 \Delta$ then we are forced to go back to adding probability amplitudes rather than probabilities themselves. 

The only way to reconcile the two is, of course, decoherence (see [13]). If there is enough random processes going on between the emission and the detection of electrons, the interference pattern will become chaotic and average to $0$, which means that then due to random phase shifts the probabilities predicted by sums of amplitude will coincide with the sum of actual probabilities. In order for this decoherence phenomenon to ALWAYS happen whenever we are dealing with scales larger than $\Delta$,  we have to have some lower bound on "entropy density" so to speak so that on the scale $\Delta$ we always have enough of accumulated entropy for the decoherence to occur. This, of course, is equivalent to the third law of thermodynamics which, based on empirical observations, demands a lower bound of absolute temperature. According to this paper, these "empirical observations" are explained by the preclusion of the universes where they don't hold. Or, more precisely, it is possible to have non-precluded universe with absolute $0$ temperature, but in that case we have to "preclude" any possible experiments (including double slit one) that would pose a problem -- in other words, we would have some trivial universes, such as pure vacuum.

Now lets go back to the scales smaller than $\Delta$. An important question to ask is this: suppose we have a complex system that lives inside a region much smaller than $\Delta$; will the decoherence effects produce measurement phenomenon still occur, despite the fact that we have only explicitly postulated that they would on scales greater than $\Delta$ ?  The answer to this, which is vital to the validity of the theory, is yes. Suppose a particule interacts with some complex system between $t=A$ and $t=B$ where $B-A < \Delta$. Consider a third point, $t=C=B+ \Delta$. Since both $C-A$ and $C-B$ are greater than $\Delta$, we would expect appropriate quantum correlations both between A and C and between B and C. The "appropriate quantum correlation between A and C" would be defined based on Schrodinger's equation on that interval. The crucial step in the argument is that Schrodinger's equation will include decoherence resulting in interaction with a complex system between A and B. This means that, on the one hand, we are NOT asked to make sure that A and B correlates in appropriate way; but AT THE SAME TIME we are asked to do that between B and C we will "change our memories" in a way that at point C we "remember" that the appropriate correlation between A and B did occur. However, due to the fact that $C-B > \Delta$, we can not do anything "illegal" between B and C; in particular, we are not allowed to "change memories". This means that we better do have appropriate correlation between A and B, despite the fact that we were not demanded to do that by constraint. The only objection to this argument is that by going "backward in time" from C to B one can see that different scenarios at B might correspond to the same "memory" at C. The answer to this objection is that by considering a lot of different values of $C$, our restrictions for B will be stricter and stricter to meet the constraints for all possible $C$-s at the same time. While this is a guess work rather than a theorem, qualitatively it is very plausible that by doing that we would effectively restrict $t=B$ to what is expected on the $(A,B)$ interval, despite it never being an official constraint.

It is important to note that the "memory" mechanism is decoherence; thus the above argument only implies to the situation where the interaction with complex system occurs between A and B. This means that we can have it both ways: on the one hand, we avoid Zeno effect if we don't have complex system, and on the other hand we claim that if we do have a measurement apparatus in a form of complex system, we will produce both quantum measurement phenomenon as well as Zeno effect on arbitrarily small time scales, as long as they are large enough for us to "fit" that complex system into them. The only difference between the prediction of this theory is that if there are no complex systems, we would still have a "localization" on time scales greater than $\Delta$. However, since, as we said earlier, $\Delta$ can be very large, we can always dismiss any experimental evidence of quantum effects occuring on astronomical scales by claiming that $\Delta$ is even larger. 
 
It is important to note that this approach is "torn" between Bohm's and Everett's views. Bohm claims that particle moves along a single trajectory, which he refers to as position beable, while Everett claims that we have a wave function alone and each of its peaks which are produced by decoherence represent parallel universe. In this paper, we are sitting between two chairs: on the one hand we have postulated that particle behavior is well defined at every instance of time, which is Bohm-like; on the other hand, we never imposed its equation of motion which Bohm did, and instead replace it with preclusion, which in some sense might be seen as Everett-like. This might also be viewed as "incomplete" Bohm's model: we can interpret Bohm's guidence equation as a "preclusion" of every single trajectory except for one; we, on the other hand, left more than one trajectory un-precluded. This raises a question: can we retain all of the conclusions of Bohm's model despite being "incomplete"?  

 According to Bohm's Pilot Wave model, a particle evolves according to guidence equation, which roughly speaking means that it moves in the direction of gradient of a wave function. If the decoherence of the wave function occurs, it will split into several branches, and the particle will end up being in one of them. Since, according to decoherence theory, the branches will not overlap in the future, the particle will stay inside one of the branches; this is the consequence of the fact that its direction of motion is parallel to the gradient of the wave function. Since in our case we no longer postulate an equation to this amount, our model makes much weaker statement regarding the behavior of a beable than Pilot Wave model does. So what we would like to ask ourselves is whether or not there is enough of an overlap to still retain the conclusion. 

On the example of a dice, if we throw it infinitely many times, we have 100 percent probability that at least once we would get the same side 100 times in a row, which would by itself be considered as low probability scenario. From this we see that low probability scenarios are allowed, as long as they occur limited number of times. From this point of view, it is possible that a particle would jump from one of the branches to the other, as long as it doesn't do it too often. However, it is important to note that if a particle jumps from branch A to branch B and then stays in branch B, then it would be a lot more than one low probability event. After all, we can choose between many different instances in time when the particle resided in branch A, and we can likewise choose between many instances when it resided in branch B, and no matter which choice we make, we would get a new example of low-probability-correlation. In fact, if we imagine that a particle stayed in both of these branches for equal periods of time, then fifty percent of choices of pairs of  events will come from two different branches. Clearly, if 50 percent of events have low probability, the whole history will be forbidden. On the other hand, if we consider a history that a particle jumps from branch A to brach B and then returns to branch A right away, then as long as the period of time it resides in branch B is short enough, most of the pairs of events will be coming from the same branch, branch A, thus the whole history would have high enough probability in order to be one of the allowed ones.

Thus, the difference between outcomes of Pilot Wave model and ours is that according to Pilot Wave moded the particle is dynamically dictated to stay inside of the branch it happened to get into, while according to our model the particle can jump back and forth between branches, as long as there is one branch in which it stays most of the time. Since, by nature, any classical observation neglects microscopic process, "most of the time" is more than enough as far as classical phenomena are concerned. Thus, we can take that weaker statement and still imply that decoherence effects does occur, just like we were hoping.  

\bigskip

\noindent{\bf 3. Relativistic bosonic field in fixed gravitational background}

Throughout this section we will use the following definitions:

DEFINITION: Let $p$ and $q$ be two elements of Lorentzian manifold $M$. We say that $p \prec q$ , and also that "p is in a causal past of q" if and only if you can travel from $p$ to $q$ without going faster than the speed of light. Whenever either $a \prec b$ or $b \prec a$ we say that $a$ and $b$ are causally related. 

DEFINITION: Let $S_1$ and $S_2$ be two subsets of $M$. We say that $S_1 \prec S_2$ if we can not find $a \in S_1$ and $b \in S_2$ such that $b \prec a$. At the same time, there exist at least one pair of points $c \in S_1$ and $d \in S_2$ such that $c \prec d$

DEFINITION: Let $p$ and $q$ be two points on a manifold $M$. In this case $\tau(p,q)$ is defined to be the length of the shortest geodesic $\gamma$ that connects $p$ and $q$, which is given by

\bea \int_{\gamma} \sqrt{\vert g_{\mu\nu}dx^{\mu}dx^{\nu}\vert} \eea

We would now generalize the approach described in the previous section for the case of space time with fixed curvature. Naively, we could have simply replaced points in time with points in space-time. However, this approach does not work. Consider, for example, a strong positive charge that attracts two negative charges to it. If we single out the location of the two negative charges, we would get an un-probable event: the two negative charges are close together without any positive charge that holds them. This means that we have to be careful to take into account all "relevent information". Relativistic invariance, however, provides a guidence for us as to what the "relevent information" is, namely it is the past light cone. However, if we were to consider the entire light cone, we would have been able to come up with scenarios that yesterday has wrong correlation with today, which is okay because the day before yesterday "fixed" it. It is true that on the example in the previous section we did just that when we confirmed that it is possible for a particle to jump from branch A to branch B of the decoherence pattern as long as it returns to branch A right away. But at the same time we also remarked that this should not be happening consistently. But if we are to take the entire light cone into account, we might as well have a CONSISTENT pattern of two conseqitive days not matching each other because of some other event in the past. The way to adress this issue is to cut a light cone with a surface. If we only had one event, then defining the surface in terms of Lorentzian distance to that eventwould imply that the surface were hyperbolically approaching light cone, giving us some unpleasant singularities. But since a lot of quantum mechanics problems are dealing with systems of size more than one point, we might assume that such is always the case which allows us to view a single event as a two different events that are spaced very closely to each other. In this case, we might define a surface in terms of a minimum of the Lorentzian distances to these two events. It is easy to see that that surface is compact. We will now generalize it to the case when the number of these points can be both finite and infinite. We will call the set of these points $S$: 
\bea \nonumber \eea
DEFINITION: Let $M$ be a Lorentzian manifold, and let $T$ be some set of events on that manifold. The past shaddow of $T$ of order $\tau$ is defined to be 

\bea S_{past}(T,\tau) = \{ p \in M \vert sup \{ \tau(p,q) \vert q \in T \; and \; p \prec q \}  = \tau \} \eea

Likewise, a future shaddow of $T$ of order $\tau$ is defined to be 

\bea S_{future}(T, \tau) = \{ p \in M \vert sup \{ \tau(p,q) \vert q \in T \; and \; p \succ q\}  = \tau \} \eea

I will now define the probability amplitude of the transition from field configuration on the earlier surface $L_1$ to the field configuration on the later surface $L_2$ to be  

\bea {\cal A} ({\cal F} ; T_1, T_2) = \frac{\int_{{\cal F} (T_1) = {\cal G} (T_1) ; {\cal F} (T_2) = {\cal G} (T_2) } [d {\cal G} ] e^{i S ({\cal G})} }{\int_{{\cal F} (T_1) = {\cal G} (T_1) } [d {\cal G} ] e^{i S ({\cal G})} } \eea  

This definition matches what is standardly done in quantum field theory. When we compute a set of propagators, $\frac{\partial^n}{\partial J(p_1) . . . \partial J(p_n)} \int [d {\cal F}]e^{S({\cal F}, {\cal J})}$ for some set of fields ${\cal F}$ and some set of currents ${\cal J}$, we are implicitly assuming that ${\cal F}$ and ${\cal J}$ are two separate interacting fields. We then compute probability amplitude associated with specific values of ${\cal J}$, namely that field having non-zero values at certain fixed points and zero values everywhere else. To do that, we integrate over everything that is non-defined, namely ${\cal F}$. Since source field is just another field, there is no reason for it to receive a "beneficial treatment" when it comes to general interpretation of quantum mechanics. At the same time, if we treat all fields as "source fields" there would be no "dummy indices" left to integrate over; on the other hand, if none of the fields are sources, then everything is dummy index hence there is no meaningful information left. So instead of limitting the "observables" to one specific set of fields, we decide to limit observables to specific set of points; in that set of points, all fields are source fields, while everywhere else all fields are dummy indices. Since shortly, mimicking the procedure done in the previous section, we will be considering all possible choices of such sets of points, our overall procedure treats all points on equal footing. This is different from what was done with source currents since in that case, we assume source field is defined everywhere -- in particular, outside of the select points it is assumed to be $0$. On the other hand, in our case we are no longer assuming that fields are $0$ outside of the fixed set of points. That is what allows us to treat all fields as "sources" while still having degrees of freedom to integrate over. 
 
The probability density will be  

\bea p_A ({\cal F} ; T_1, T_2 ) = \vert {\cal A} ({\cal F}; T_1, T_2 ) \vert ^2 \eea

Here, we call it $p_A$ because we will later introduce $p_B$ and $p_C$ as alternative ways of defining probability. Finally, we would like to impose a constraint on the global correlation which is spacetime analogue of constraint imposed in the previous section:

CONSTRAINT 1 VERSION A: Let $M$ be a d-dimensional manifold. For every number $n$, if we have points $q_1, . . . , q_n$,

\bea \forall \tau > \Delta \big[p_0- \epsilon_1< \mu \{q_1, . . . , q_n \vert p_0-\epsilon_2< p({\cal F}; S_{past}(\{ q_1 , . . . , q_n \}, \prec, \tau), \{ q_1, . . . , q_n \} ) < p_0 + \epsilon_2 \} < \nonumber \eea
\bea < p_0 + \epsilon_1 \big]  \eea

This, however, might prove to be too restrictive given that Lorentz group is non-compact, so there are infinitely many reference frames we have to simultaneously satisfy, where a reference frame roughly corresponds to choice of points $q_1$ through $q_n$ . Since I don't know for sure one way or the other in terms of whether it is possible to simultaneously satisfy all of these constraints, I will write an "easier" version of the above constraint, version b. So, if in the future work it turns out that "harder" version (version a) is too restrictive, easier version will be used; or if it turns out that harder version is not too restrictive, then harder one will be used. 

In designing the easier version I will appeal to the following observation: while from the fundamental physics point of view all frames are equivalent, on practice in any given region of space-time fields are propagating with limitted velocity relative to each other. This is reflected in the following observation: on the one hand, fields are locally smooth, while on the other hand fields can vary by arbitrary large amount in the vicinity of light cone despite the fact that Lorentzian distance is arbitrary small in that region. The flip side of a coin of this argument is this: suppose we have a wave packet whose shape doesn't change in its own reference frame (this, of course, never happens due to the nature of wave equation, but this is good enough for a simple toy model to illustrate my point). Then in the frame moving arbitrarily close to the speed of light with respect to wave packet, wave packet will undergo Lorentz contraction to an arbitrarily small size; this means that as a wave packet passes a given point, the fields change arbitrarily fast. Due to the fact that Lorentz group is non-compact, most of the wave packets should be moving arbitrarily close to the speed of light with respect to any given frame. This means that fields should fluctuate arbitrarily fast in any given frame. The fact that this is not happening implies that the fields "picked" a frame with respect to which not to move too close to the speed of light. 

Even though this sounds like violation of relativity, based on what we have just said, it is simple consequence of fields having bounded derivative. Thus, it can be enforced with the following, Lorentz covariant, constraint:

CONSTRAINT 2: $\partial^{\mu} \partial_{\mu} \phi$, $(\partial^{\mu} V^{\nu})(\partial_{\mu} V_{\nu})$ as well as any other contractions of partial derivative with itself are all bounded by some fixed large number, $D$.

Now, going back of designing an easier version of constraint 1, we will do the following: we will let ourselves "shift" a shaddow by a very small amount, in the $t$ direction as defined by a reference frame of that shaddow. In particular, we will let our displacements vary in different parts of the shaddow, as long as they are all bounded by some small number $\delta$. We will only consider the cases when the probability stays in the specified range regardless of the shift, while discarding all other choices of points. In light of what we have said regarding fast variations in some frames, this means that we only consider the points whose common "reference frame" doesn't move too fast with respect to the "ether".  

First, let us provide some definitions

DEFINITION: Let T be a set of points and let U be some set. We say that U is "shifted past $\tau$-haddow" of T of tolerance $\delta$ if the following is true: 

a)Any element of $U$ is to the future of at least one element of $S_{past} (T, \prec, \tau)$ but is not to the past of any of the elements of the above. 

b)If $u \in U$ and $r \prec u$ for some $r \in S_{past} (T, \prec, \tau)$, then $\tau (r, u) < \delta$ where $\tau$ stands for Lorentzian distance and $\delta$ is some small constant

c)If $\gamma$ is any curve that connects an element of $S_{past} (T, \prec, \tau)$ and an element of $T$, which is always either timelike or lightlike and is directed to the future, then $\gamma$ passes exactly one element of $U$

We can likewise define a shifted future shaddow:

DEFINITION: Let T be a set of points and let U be some set. We say that U is "shifted future $\tau$-shaddow" of T with tolerance $\delta$ if the following is true: 

a)Any element of $U$ is to the past of at least one element of $S_{future} (T, \prec, \tau)$ but is not to the future of any of the elements of the above. 

b)If $u \in U$ and $u \prec r$ for some $r \in S_{future} (T, \prec, \tau)$, then $\tau (r, u) < \delta$ where $\tau$ stands for Lorentzian distance and $\delta$ is some small constant

c)If $\gamma$ is any curve that connects an element of $S_{future} (T, \prec, \tau)$ and an element of $T$, which is always either timelike or lightlike and is directed to the past, then $\gamma$ passes exactly one element of $U$

Finally, we write an "easier" version of constraint 1:

CONSTRAINT 1 VERSION B: Let $M$ be a d-dimensional manifold. For every number $n$, if we have points $q_1, . . . , q_n$,

\bea \forall \tau > \Delta \big[p_0- \epsilon_1< \mu \{q_1, . . . , q_n \vert \; if \; T \; is \; \delta \; shift \; of \; S_{past}(\{ q_1 , . . . , q_n \}, \prec, \tau) \; \nonumber \eea
\bea then \; p_0-\epsilon_2< p({\cal F}; T, \{ q_1, . . . , q_n \} ) < p_0 + \epsilon_2 \} < p_0 + \epsilon_1 \big]  \eea

As was previously said, the implication of our ability to select a "shifted" shaddow is that if fields fluctuate a lot, then the variations of $p$ as we shift a shaddow would be too large to fit into any specified range, hence this would not provide any information of interest. Thus, the above constraint only applies to reference frame of small variation. 

Finally, we will show an alternative procedure to "version B" that would have the same affect: a floating lattice. This we will call version C. In this approach we note that in order for path integral to be rigorously defined, we need to employ some kind of discretization anyway. So we will take advantage of that and use the discritized shaddow surface as a replacement for continuum based "shifted" shaddow we just introduced. The key factor in both cases is that neither "shifted" shaddow nor "discritized" shaddow coincides with original shaddow, but both can be regarded as its approximation in the case of fields not fluctuating too fast in reference frame defined by the shaddow. Hence, if we ignore the issue of infinitely many degrees of freedom, we can claim that if we have one of the two we don't need the other. The advantage of "discritized" shaddow option is the fact that we do have to remember the infinity issue at some point. On the other hand, the advantage of "shifted" shaddow is to make a point that the above approach, in its spirit, is continuum based and infinity issues are un-related to it. So to meet both agendas I will keep both approaches as far as this paper is concernted.

DEFINITION: Let $T$ be a set of points. Its past $\tau$-submanifold is a union of its past $\sigma$-shaddows corresponding to all $\sigma \leq \tau$ 

DEFINITION: Let $T$, $U$, and $V$ be some finite sets of points, and suppose $U \subset V$. We say that $U$ is discritized past $\tau$-shadow of $T$ and $V$ is discritized past $\tau$-lattice based on $T$ if  the following statements are true: 

a)Every element of $V$ is causally after at least one element of the $\tau$ - past shaddow of $T$ and causally before at least one element of $T$

b)Every single element of past $\tau$-shaddow of $T$ is to the past of at least one element of $V$. Likewise, every single element of $T$ is to the future of at least one element of $V$

c)If the volume of past $\tau$-submanifold of $T$ is $I$, then the number of elements of $V$ is less than $I/v_1$ and greater than $I/v_2$

d)If $p$ and $q$ are two elements of $V$ and Lorentzian distance between them is greater than $\tau_0$ then there are at least $n$ points $r$ that are to the future of $p$ and to the past of $q$ (where my definition of past and future includes the fact that they are timelike-separated from $p$ and $q$ )

e)If $r \in U$ then there exist $s$ that is to the causal past of $r$ such that the Lorentzian distance between $r$ and $s$ is less than $\tau_0$ but at the same time we have MORE than $m$ points that are part of the shaddow of $T$ and are both to the causal future of $s$ and to the causal past of $r$. This criteria assures that the reference frame defined by discritized shaddow approximates the one defined by actual shaddow and rules out the issues of lightcone singularities.

We will likewise introduce a definition of "future" discritized shaddow: 

DEFINITION: Let $T$ be a set of points. Its future $\tau$-submanifold is a union of its future $\sigma$-shaddows corresponding to all $\sigma \leq \tau$ 

DEFINITION: Let $T$, $U$, and $V$ be some finite sets of points, and suppose $U \subset V$. We say that $U$ is discritized future $\tau$-shadow of $T$ and $V$ is discritized future $\tau$-lattice based on $T$ if  the following statements are true: 

a)Every element of $V$ is causally before at least one element of the $\tau$ - future shaddow of $T$ and causally after at least one element of of $T$

b)Every single element of future $\tau$-shaddow of $T$ is to the causal future of at least one element of $V$. Likewise, every single element of $T$ is to the causal past of at least one element of $V$

c)If the volume of future $\tau$-submanifold of $T$ is $I$, then the number of elements of $V$ is less than $I/v_1$ and greater than $I/v_2$

d)If $p$ and $q$ are two elements of $V$ and Lorentzian distance between them is greater than $\tau_0$ then there are at least $n$ points $r$ that are to the future of $p$ and to the past of $q$ (where my definition of past and future includes the fact that they are timelike-separated from $p$ and $q$ )

e)If $r \in U$ then there exist $s$ that is to the causal future of $r$ such that the Lorentzian distance between $r$ and $s$ is less than $\tau_0$ but at the same time we have MORE than $m$ points that are part of the future shaddow of $T$ and are both to the causal past of $s$ and to the causal future of $r$. This criteria assures that the reference frame defined by discritized shaddow approximates the one defined by actual shaddow and rules out the issues of lightcone singularities.

We will now define another "modified" version of probability. Since we are employing discreteness anyway, in the computation of probability amplitude, we will be using only the degrees of freedom associated with elements of $V$ rather than the entire past submanifold. This will allow us to avoid infinity problem. We will replace vector fields with functions $V \times V \rightarrow \mathbb{R}$ which correspond to path integrals of the vector fields along geodesics connecting relevent pairs of points. Gravity will be replaced by the partial ordering that will determine causal relations between these points. In papers [6], [7], [8] and [10] we have explored the way of rewriting Lagrangian in terms of the above coordinate-free quantities. 

Since vector fields are now functions of pairs of points, we can't strictly speaking make restriction that the "fluctuated" version of a vector field agrees with original one "only" at a discritized shaddow. Rather, our restriction should be that the two-point functions should agree for any given pair of points if AT LEAST ONE of the elements of that pair is at the discritized shaddow. 

DEFINITION: Let $S$ and $T$ be finite subsets of Lorentzian manifold and suppose $T \subset S$. If $a \colon S \times S \rightarrow \mathbb{R}$ is some real valued function on the set of pairs of points of $S$, then $A (T, \prec)$ is a set of all other functions $b \colon S \times S \rightarrow \mathbb{R}$ such that for all $p \in T$ , $a(p,q)=b(p,q)$  and $a(q,p)=b(q,p)$ for any point $q \in S$, regardless whether $q$ is an element of $T$ or not.

We then define the transition amplitude in the following way: 

\bea {\cal A}_{discrete} (\phi_1, . . . , \phi_n , a_1 , . . . , a_m ; T_1, T_2, S) = \nonumber \eea
\bea = \Big(\sum_{ \{ a_1^* \colon S \times S \rightarrow \mathbb{R}, . . ., a_m^* \colon S \times S \rightarrow \mathbb{R}, \phi_1, . . . , \phi_n \vert a_k^* \in A (T_1 , \prec) \; \} } e^{i S ( \phi_1, . . . , \phi_n, a_1, . . . , a_m)}\Big)^{-1} \times \nonumber \eea
\bea \times \Big(\sum_{ \{ a_1^* \colon S \times S \rightarrow \mathbb{R}, . . ., a_m^* \colon S \times S \rightarrow \mathbb{R}, \phi_1, . . . , \phi_n \vert a_k^* \in A (T_1 \cup T_2, \prec) \; and \; }  e^{i S (\prec, \phi_1, . . . , \phi_n, a_1, . . . , a_m)}\Big) \eea

and 

\bea p_{discrete} (\phi_1, . . . , \phi_n , a_1 , . . . , a_m ; T_1, T_2, S) = \vert {\cal A} (\phi_1, . . . , \phi_n , a_1 , . . . , a_m ; T_1, T_2, S) \vert^2 \eea

We  now define our "version C" of the probability:

CONSTRAINT 1 VERSION C: Let $M$ be a d-dimensional manifold. For every number $n$, if we have points $q_1, . . . , q_n$,

\bea \forall \tau > \Delta \big[p_0- \epsilon_1< \mu \{q_1, . . . , q_n \vert \; if \; T \; is \; \delta \; discritization \; of \; S_{past}(\{ q_1 , . . . , q_n \}, \prec, \tau) \; \nonumber \eea
\bea then \; p_0-\epsilon_2< p({\cal F}; T, \{ q_1, . . . , q_n \} ) < p_0 + \epsilon_2 \} < p_0 + \epsilon_1 \big]  \eea

\bigskip

\noindent{\bf 4. Modifications for fermionic case}

Adapting the above approach for fermionic case is tricky because if we are to assume the existance of fixed distribution of fermionic field, as was done for the bosonic case, this requires defining Grassmann numbers outside of integration. This I have already done in [9] In that paper, I have defined Grassmann numbers to be elements of a space equipped with commutting dot product ($\cdot$), anti-commutting wedge product ($\wedge$) and a measure $\xi$ that has both positive and negative values. The vector space is multidimensional with unit vectors $\hat{\theta_1}$, $\hat{\theta_2}$, etc. and these are ordered. The relation between products is $\hat{\theta_i} \cdot \hat{\theta_j} = \hat{\theta_i} \wedge \hat{\theta_j}$ for $i<j$ (which means $\hat{\theta_i} \cdot \hat{\theta_j} = - \hat{\theta_i} \wedge \hat{\theta_j}$ for $i>j$). We further define $\hat{\theta_1} \cdot \hat{\theta_1} =1$,  $\hat{\theta_1} \cdot (\hat{\theta_1} \wedge \hat{\theta_2}) = \hat{\theta_2}$, $(\hat{\theta_1} \wedge \hat{\theta_2}) \cdot (\hat{\theta_1} \wedge \hat{\theta_2})=1$, $(\hat{\theta_1} \wedge \hat{\theta_2}) \cdot (\hat{\theta_1} \wedge \hat{\theta_3}) = \hat{\theta_1} \wedge \hat{\theta_3}$, etc. We also have a measure $\xi(\theta)$ that satisfies

\bea \int \xi(\theta) d \theta = 0 \; ; \; \int \theta \xi(\theta) =1 \eea

In the previous paper, it was shown that in this case

\bea \int_{usual} d \theta_1 . . . d \theta_n \xi(\theta_1) . . . \xi (\theta_n) (\hat{\theta_1} \wedge . . . \wedge \hat{\theta_n}) \cdot f(\theta_1 \hat{\theta_1}, . . . , \theta_n \hat{\theta_n}) = \nonumber \eea
\bea =\int_{Grassmann} d \theta_1 . . . d \theta_n f(\theta_1 , . . . , \theta_n) \eea 

as long as on the left hand side $f$ is expressed in terms of wedge products. 

Furthermore, in that paper we defined a spinor field in a more geometric way. In particular, we have considered two scalar fields, $\chi_p$ and $\chi_a$ (one for particle and one for antiparticle) and a four non-orthogonal vierbines, $e_1^{\mu}$, $e_2^{\mu}$, $e_3^{\mu}$ and $e_4^{\mu}$. We assume that in the reference frame defined by these non-orthonormal vierbines, our spinor field is given by $(\chi_p, 0, \chi_a, 0)$ From these four non-orthonormal vierbines, we use Gramm Schmidt process to derive the four orthonormal ones: $f_a^{\mu} = f_a^{\mu} (e_1, e_2 , e_3 , e_4)$ We then rotate a spinor $(\chi_p, 0, \chi_a, 0)$ from the u-based coordinate system to v-based one. This will give us a complete spinor whose all four components are non-zero. Thus, we have a function $g^s$ that takes four non-orthonormal vierbines and two real scalar fields and returns a spinor field. We will then define a measure on the domain of $g^s$ in the following way:

\bea \lambda (\chi_p , \chi_a , e_1 , e_2 , e_3 , e_4 ) = \nonumber \eea
\bea =(\prod \frac{\partial g^s}{\partial e_k})^{-1} (\prod \frac{\partial g^s}{\partial \chi_p})^{-1} (\prod \frac{\partial g^s}{\partial \chi_a})^{-1} \prod \xi (g^s (\chi_p, \chi_a, e_1 , e_2 , e_3 , e_4 )) \eea

After having done these modifications, we will then do the following replacement:

\bea \int d \psi (p) h(\psi(p) , . . . ) \rightarrow \int (\prod d e_k^{\mu} (x)) d \chi_p (x) d \chi_a (x) \times \nonumber \eea
\bea \times \lambda (\chi_p, \chi_a, e_1 , e_2 , e_3 , e_4) (\hat{\psi_1} \wedge \hat{\psi_2} \wedge \hat{\psi_3} \wedge \hat{\psi_4}) \cdot \nonumber \eea
\bea \cdot h(\psi_1 (\chi_p, \chi_a, e_1 , e_2, e_3, e_4), \psi_2 (. . .), \psi_3 (. . . ) , \psi_4 (. . .), . . . ) \eea

Here, $\hat{\psi_s}$ are just constant vectors; they are NOT unit vectors corresponding to any variable. Thus, they do NOT imply that we are going back to the view of $\psi_s$ being independent variables.

From this point on we can simply copy everything that was done in the previous section, taking these modifications into account.

However there is one more, unrelated, thing that we might want to change for fermions. As we know from Bohm's model, while field beables are more convenient to use for bosons, position beables are more convenient to use for fermions. Among them is the fact that fermions, rather than bosons, are the main ingredients of molecules that make up the objects that are localized in space. Personally, I have not decided for sure whether I would like to turn field beables into position beables when it comes to fermions. But just in case I would like to do that, I can potentially impose a constraint which demands that, while my field is well defined at every single point in space, it is close to $0$ outside of a vicinity of some set of curves. 

CONSTRAINT 3: We still have fields $\chi_p$, $\chi_a$ and $e_a^{\mu}$ throughout space time. But the set in which $\chi_p$ and $\chi_a$ is non-zero looks like a set of piecewise continuous curves and piecewise differentiable curves whose number is no greater than $N$. While these curves are timelike at every point, they go both forward and backward in time. If I will orient them in a particular way, these curves have the following property: each of these curves starts at a point that is not to the future of any other point, and ends at a point that is not to the past of any other point (in other words, it starts at a "past" surface of space-time and ends at a "future" point of space-time) The total length of each curve is bounded by $\tau_{max}$ (by making $\tau_{\max}$ greater than the separation of past and future surfaces of the universe, but not too much greater, I will assure that most of each curve is taken up by future directed pieces). On the future directed piece of each curve, $\chi_a=0$ while on the past directed piece of each curve $\chi_p=0$. The future directed pieces are interpreted as particles, while the past directed pieces are interpreted as antiparticles. Their junctions are interpreted as pair creation and pair annihilation. 

There is no curve that represents a photon that produces pair creation or is produced by pair annihilation. We view both electrons and photons similarly to classical physics: electrons are localized while electromagnetic field that they produce fills the entire space. This agrees with the traditional approach to Pilot Wave model, where position beables are used for fermions while field beables are used for bosons. However, while pair creation or annihilation doesn't have photonic lines attached to it, there has to be a correlation between the strength of electromagnetic field and the frequency of these creation/annihilation events, which I hope will be produced once the model is investigated numerically. 

PLEASE NOTE: the curves that were given are NOT aforegiven. Rather, I am saying that a given field configuration is allowed if we can FIND a set of curves FOR THAT SPECIFIC FIELD CONFIGURATION. In other words, there are no apriori structure of a manifold in a form of curves, so no symmetries, have been violated. 

It is easy to see that if both $\chi_p$ and $\chi_a$ are small, then all four components of a vector produced by rotation of $(\chi_p, 0 , \chi_a , 0)$ from non-orthonormal to orthonormal frame will still be small, in other words, $g_s (\chi_p, \chi_a, u_k^{\mu})$ will be small regardless of values of $e_k^{\mu}$ . 

Thus, what we see above is that we are making sure that the spinor field is small, unless we are close to at least one pair of points on a segment of a curve. The reason we need more than one point is that due to the fact that Lorentzian distance is $0$ on the vicinity of a light cone, the whole space-time will meet a criteria of "close enough" if it was based on a single point. For that same reason, we impose a constraint that the Lorentzian distance between these pairs of points should be greater or equal to $\delta$. By imposing two separate inequalities for $\chi_p$ and $\chi_a$ we have also made sure that $\chi_p$ is small away from vicinity of future-directed segments while $\chi_a$ is small away from the vicinity of past-directed segments.

Two things are worth noting:

1) While $\chi_p$ and $\chi_a$ are $0$ outside of a specific set of curves, $u_a^{\mu}$ can be anything we like throughout the whole space. This means that we can use $e_a^{\mu}$ to answer a question as to how come the space attains manifold structure, and we can do so while still believing fermions are localized in space. We will talk more about it in section 6 on causal set theory.

2) The fact that we have constrained our curves to start at the point that is not causally after any other point and to end at a point that is not causally before any other point, we essentially made sure that more than 50 percent of each curve is taken up by a particle region rather than antiparticle one. Furthermore, by limitting the length of each curve by $\tau_{\max}$ we found a way of making the ratio of antiparticle part of each curve to particle one as small as we like by making $\tau_{\max}$ sufficiently small. This can be an explanation as to why our universe is matter dominated. 

\bigskip

\noindent{\bf 5. Possible validity of Dison's view that gravity does not have to be quantized}

Let us now see how can I apply my model of quantum mechanics to support the Dyson's view (see [2]) that since gravitons were never observed, we don't have to quantize gravity. The alternative of quantizing gravity will be explored in the next section.

We will start from the bosonic case when we won't have to worry about black hole singularities produced by point particles, and then we will move onto fermionic case.

As we seen in the above sections, we have a set of parallel universes, in each of them we have a manifold with well defined metric, on which all fields, both fermionic and bosonic, are likewise well defined. We then imposed some relativistically invariant, albeit global, constraints that determined whether each particular parallel universe is "allowed" or "forbidden". We then claimed that if we are living in one of the "allowed" parallel universes, we would observe the expected quantum effects.

Now, since in each parallel universe all fields are well defined, we might as well impose another constraint on whether or not a universe is "allowed" or not. Namely, that Einstein's equation is approximately satisfied. We can't say that Einstein's equation is satisfied exactly. After all, by Bianchi identities,

\bea \partial^{\mu} (R_{\mu \nu} - \frac{1}{2} R g_{\mu \nu}) = 0 \eea

but at the same time, due to the quantum fluctuations

\bea \partial^{\mu}T_{\mu \nu} \neq 0 \eea

But due to the fact that quantum processes are small, and thus have small gravitational field, we can still say that Einstein's equation holds approximately. By approximately I mean the following:

\bea R_{\mu \nu} - \frac{1}{2} R g_{\mu \nu}  = T_{\mu \nu} + t_{\mu \nu} \eea

where $t_{\mu \nu}$ is a small perturbation to $T_{\mu \nu}$, where by "small" I mean that it satisfies constraints

 \bea T_{\mu \nu}t^{\mu \nu}< \epsilon_1 \; ; \; g_{\mu \nu}t^{\mu \nu} < \epsilon_2 \eea

Thus, by making $\epsilon_1$ and $\epsilon_2$ to be of a magnitude of gravitational fields of classical objects, albeit very small, we will assure that there is no local correlation between fluctuation of gravitational field in microscopic level and any of the microscopic processes that occur. In other words, quantum mechanical particles don't gravitate, which supports Dyson's view that there is no such thing as quantum gravity, or graviton. 

This raises another question: what if we have a large body of mass which explodes and particles fly out in many different directions very far away from each other. Then, each individual particule, due to being very far from all the other particles, won't gravitate. At the same time, if their final distribution is regarded to be spherically symmetric, we can draw a very large sphere that encompasses all of these particles, and we would expect gravitational field on that sphere to be described by Schwartschild solution. This is something we can't deny since that can be potentially supported by experiments, albeit being hard to arrange. To answer this question lets introduce some names. We will call $T_{\mu \nu}$ actual energy momentum tensor, and we will call $R_{\mu \nu}- \frac{1}{2}Rg_{\mu \nu}$ an "apparent energy momentum tensor". Thus, by Bianchi identity, apparent energy momentum tensor is conserved while actual one is not. Now, back at the time when the mass have not spread out yet, we agree that due to its density being sufficiently large, both apparent and actual energy momentum tensors were large, and they were approximately equal to each other. Then, after the mass spread out, the actual energy momentum tensor became very small, which means that the apparent one no longer needs to have any local correlation with an actual one. However, due to the fact that apparent energy momentum tensor USED to approximate an actual one when they both WERE large, the conservation of apparent energy momentum tensor demands that it should look like some kind of distribution of the latter throughout space. True, that distribution doesn't have to correlate with specific behavior of particles that are being spread out. But since we can not measure the behavior of these particles, as long as this distribution obeys the conservation law, there is no way we can "get caught" with violation of general relativity. 

Now one thing to adress is how do we avoid black hole singularity of gravitational field produced by fermions, if we take a view that they are localized into curves. The way I avoid this is that when I write $T_{\mu \nu}$ for fermionic field, I replace $\chi_p$ and $\chi_a$ with $\chi '_p$ and $\chi '_a$ which are "blured out" version of original fields, that is, they are non zero in a close vicinity of a curve. They are defined as follows:

DEFINITION: Let $r$ be a point on a manifold $M$. Let $A_p$ (index $p$ stands for particle) be a set of pairs of points $(p,q)$ such that they both fall on the same future-directed piece of a curve, and are also both lightlike separated from $r$. If $A_p (r)$ contains no elements, then $\psi '_p (r) =0$. On the other hand, if it does contain some elements, then we let $A_{min;p} (r)$ to be a subset of $A_p (r)$ that minimizes Lorentzian distance between the pairs of elements (so with probability 1 it will be one element set but I just do it for the sake of generality). Let $\tau_{min}$ be the Lorentzian distance between two elements of any of the pair belonging to $A_{min;p} (r)$. Then 

\bea \chi '_p (r) = \rho (A_{min;p} (r)) \frac{ \sum_{(p,q) \in A_{p; min} (r)} (\chi_p (p) + \chi_p (q))}{2\sharp A_{p; min} (r)} \eea

, where $\rho(x)$ is some function whose value is $0$ outside of vicinity of $0$. 

The reason we define the vicinity of a curve in terms of two points rather than one point is that due to the fact that Lorentzian distance is $0$ in the lightcone of one point, we would have been able to fill the whole space. On the other hand, by considering two points that are defined by intersection of lightcone comming from $r$ with a curve representing a path of a fermion, we make sure that point $r$ is very close to a curve, in a reference frame defined by that curve, which is what we want.

We will now do the same definition for anti-particle. This we do by cut and paste of definition above, while replacing $A_p$ with $A_a$, "future-directed segment" with "past-directed segment", $\psi_p$ with $\psi_a$ and $\psi '_p$ with $\psi '_a$ :

DEFINITION: Let $r$ be a point on a manifold $M$. Let $A_a$ (index $a$ stands for particle) be a set of pairs of points $(p,q)$ such that they both fall on the same future-directed piece of a curve, and are also both lightlike separated from $r$. If $A_a (r)$ contains no elements, then $\psi '_a (r) =0$. On the other hand, if it does contain some elements, then we let $A_{min;a} (r)$ to be a subset of $A_a (r)$ that minimizes Lorentzian distance between the pairs of elements (so with probability 1 it will be one element set but I just do it for the sake of generality). Let $\tau_{min}$ be the Lorentzian distance between two elements of any of the pair belonging to $A_{min;a} (r)$. Then 

\bea \chi '_a (r) = \rho (A_{min;a} (r)) \frac{ \sum_{(p,q) \in A_{a; min} (r)} (\chi_a (p) + \chi_a (q))}{2\sharp A_{a; min} (r)} \eea 

As far as the vierbines, they will be left un-changed, since I have no reason to expect them to be singular in vicinity of the curve:

\bea e'_{k \mu} (r) = e_{k \mu} (r) \eea

The important thing to notice is that fermionic action should be based on $\chi$ rather than $\chi'$ and AT THE SAME TIME $\chi'$ rather than $\chi$ for gravity. If we were to use $\chi'$ for fermionic action we would likely run on a famious problem from classical physics as to why doesn't the electron explode due to same-charge repulsion. At the same time, if we were to use $\chi$ for gravity, we would end up with a black hole. But by using $\chi$ for fermionic action and $\chi'$ for gravity, we have basically said that while the gravitating matter of a fermion is distributed over some small region, its charge is all concentrated at the center (the latter being pointlike). This means that neither same charge repulsion nor black hole happens. 

\noindent {\bf. 6 Possible quantization of gravity through causal set approach}

We will now explore an alternative where we do quantize gravitational field. While there are a lot of different models for quantum gravity, and it is certainly worthwhile to explore the possible implications of our approach to each one of them, for the purposes of this paper we will limit ourselves to causal set theory. The reason I made this choice is because causal set is a discritized space-time whose microscopic structure is manifestly Lorentz invariant outside of path integration. Since this paper is built on the concept of existance of well defined field beables at every point that are relativistically invariant, if we are to discritize our space-time at all, we better do so without violating the relativistic invariance that we strive to preserve. This makes causal set approach is an ideal one. 

A causal set, which was first introduced by Rafael Sorkin (for reviews see [3] and [4]) is any set $S$ with partial ordering $\prec$, where we interpret its elements as physical events, while the partial ordering $\prec$ we interpet as causal relations between these events. If $a \prec b$ this means that $a$ is in a causal past of $b$. That set is assumed to be discrete with respect to partial ordering. That is, if $a \prec b$ then there are only finitely many elements $c$ that satisfy $a \prec c \prec b$. No other structure, including coordinate system, are assumed. This was motivated by the observation made by Hawking where he have shown that if we have Lorentzian manifold we can describe its metric, up to Weil scaling, on the basis of causal relations alone. In the discrete scenario, Weil scaling can be defined by the simple count of points on any given region. Thus, for discrete case, causal relations will give us a complete information about metric. The motivation for a causal set is that it is a way to discritize space time without violating Lorentz covariance. If we try, for example, to discritize space time using lattice then edges of a lattice will be different from the diagonals, making one direction "better" than the other. On the other hand, in case of causal set, since the only structure is a causal relation, which is manifestly covariant, the covariance holds by definition.  

In papers [6], [7], [8] and [10] , we proposed a way to define fields and their Lagrangians for a causal set. Scalar fields are defined in a usual way, as complex valued function. Vector fields are defined in terms of real valued functions on a set of pairs of points of a causal set (in case of a manifold we interpret them as a path integral of a vector field along geodesic that connects these two points) . Gravitational field is defined in terms of causal structure itself, by appealing to Hawking's observation. And finally spinor field are described in terms of four vector fields and two scalar fields in the same way as described in previous sections of this paper as well as [9] and [10]. Since vector fields are defined in terms of real valued functions on a set of pairs of points, this unltimately implies that fermions are described in terms of combination of real valued functions on a set of single points ($\chi_p$ and $\chi_a$) together with real valued functions on a set of pairs of points (the ones corresponding to $u_k^{\mu}$. This means that if we put together scalar field, gauge field, fermionic field, and gravity, we have an action of the form

 \bea S= \sum_p {\cal L} (p; \prec, \phi_1, . . . , \phi_n , a_1 , . . . , a_m ) \eea

where $\phi_i$ are real valued functions on a set of single points and $a_j$ are real valued functions on a set of pairs of points. The actual explicit equation for ${\cal L}$ you can find in these papers. 

However, while the actions were defined, one question was not adressed: what to do with these actions? After all, since causal relations are viewed as a gravitational field, it is essentially a dummy index we are integrating over, so no afore-given causal relations are defined. Since causal relations are the only defining feature of topology, this means that any pair of elements of causal set is apriori just as far or just as close to each other as any other pair of elements, which means that their propagators, if they are defined at all, should be identical. This is where the approach used in this paper comes to rescue: we have a different parallel universes in each of which causal relations are given. Our only task is to determine which of these universes is allowed and which is forbidden. As we perform our task of determining that we, of course, will have to compute some version of propagators for fields. But we wouldn't have trouble doing that since for each particular universe we are testing, we will simply assume the causal relations of that particular universe. In other words, gravitational field will be a causal set version of relativistically covariant beable talked about previously in much the same way other fields are such. We then have to introduce fluctuations away from this beable configuration everywhere outside a given set of fixed points and their $\tau$-shaddow. 

We now have to adapt what was done before to causal set approach. First of all, when we were defining shaddow we were using Lorentzian distance. So we have to define the Lorentzian distance for causal set. We simply adapt a definition of Lorentzian distance given in [11] and [12]. First consider a flat Minkowski space and two timelike-separated points in that space. We can rotate coordinate system to make sure that these two points are lying on $t$-axis, with coordinates $t_1$ and $t_2$. If we have an arbitrary future-directed curve $\gamma$ that connects them, then its length is 

\bea l(\gamma) = \int_{\gamma} \sqrt{(dt)^2-\sum (dx_k)^2} \leq \int_{\gamma} \vert dt \vert = \int_{\gamma} dt = t_2 - t_1 = \tau (p,q) \eea

The second equal sign in above equation is based on the assumption that the curve is future-directed. Thus, while it is not true that the length of every single curve that connects $p$ and $q$ is less than the Lorentzian distance between them, it is true about future directed curves.  For example, if we didn't impose a constraint that a curve is future directed, then by going a million light years to the future and back we would of ended up with curve whose length is arbitrary large, but if we limit a curve to be future-directed, this is not possible. 

Now, in [11] and [12] they simply copied the above statement for the discrete case of causal sets. In this case, they can replace a future-directed curve with a future-directed set of points, $p \prec r_1 \prec . . . \prec r_n \prec q$. Selecting the longest possible curve can be replaced by selecting a chain of points that has largest possible number of points. Apart from the above, there is a side benefit to this: they are automatically assured that that chain of points does, in fact, approximate a curve since by making sure that it has largest possible number of points they have also made sure that they haven't "skipped" over any points, which means that the points are spaced as densely as possible. 

Numerical studies were done in [11] and [12] where it was confirmed that in case of Poisson distribution of points on a Lorentzian manifold there is, in fact, a close correlation between the Lorentzian distance between the two events and the length of the leongest chain of points that connects them. Thus, from now on I will simply assume that for a causal set a Lorentzian distance is defined in the following way:

\bea \tau (p,q) = max \{ n \vert \exists r_1, . . . , r_n \colon p \prec r_1 \prec . . . \prec r_n \prec q \} \eea

DEFINITION: Let $M$ be a Lorentzian manifold, and let $T$ be some set of events on that manifold. The past shaddow of $T$ of order $\tau$ is defined to be 

\bea S_{past}(T,\prec, \tau) = \{ p \in M \vert max \{n \vert \exists r_1 , . . . , r_n, q \colon p \prec r_1 \prec . . . \prec r_n \prec q \in T \} = \tau \} \eea

Likewise, a future shaddow of $L$ of order $\tau$ is defined to be 

\bea S_{future}(T,\prec, \tau) = \{ p \in M \vert max \{n \vert \exists r_1 , . . . , r_n, q \colon p \succ r_1 \succ . . . \succ r_n \succ q \in T \} = \tau \} \eea

The definition of "shifted" shaddow can likewise be translated into a causal set context:

DEFINITION: Let T be a set of points and let U be some set. We say that U is "shifted past $\tau$-shadow" of T of tolerance $\delta$ if the following is true: 

a)Any element of $U$ is to the future of at least one element of $S_{past} (T, \prec, \tau)$ but is not to the past of any of the elements of the above. 

b)If there is a sequence of points $r_1 \prec . . . \prec r_n$ where  $r_n \in U$ and $r_1 \in S_{past} (T, \prec, \tau)$, then we have $n \tau_0 < \delta$ where $\tau_0$ stands for shortest possible discritized distance in a causal set (i.e. a Lorentzian distance between any points $p \prec q$ for which there is no $r$ satisfying $p \prec r \prec q$ )

c)Suppose we have a sequence of points $r_1 \prec . . . \prec r_n$ where $r_1$ is an element of $S_{past} (T, \prec, \tau)$ and $r_n$ is an element of $T$. Then either $r_k \in U$ for some k or else there exist $s \in U$ satisfying $r_k \prec s \prec r_{k+1}$ for some k.

We can likewise define a shifted future shaddow:

DEFINITION: Let T be a set of points and let U be some set. We say that U is "shifted future $\tau$-shadow" of T of tolerance $\delta$ if the following is true: 

a)Any element of $U$ is to the past of at least one element of $S_{future} (T, \prec, \tau)$ but is not to the future of any of the elements of the above. 

b)If $r_1 \prec . . . \prec r_n$ is any sequence of points satisfying $r_1 \in U$ $u \in U$ and $r_n \in S_{future} (T, \prec, \tau)$, then we have $n \tau_0 < \delta$ where $\tau_0$ stands for shortest possible discritized Lorentzian distance of a causal set and $\delta$ is some small constant

c)Suppose we have a sequence of points $r_1 \prec . . . \prec r_n$ where $r_1$ is an element of T and $r_n$ is an element of $S_{future} (T, \prec, \tau)$. Then either $r_k \in U$ for some k or else there exist $s \in U$ satisfying $r_k \prec s \prec r_{k+1}$ for some k.

We have explicitly written down shadows as dependent on causal relation $\prec$ since the latter is viewed as gravitational field, thus the fact that we use gravitational field in defining what we mean by a shadow is crucial in that we are using those same shaddows in evaluation whether or not a given gravitational history is allowed or forbidden.

The other thing we have to define is restrictions of the fields to $T$ and its shaddow. The difficutly is that vector, spinor and gravitational fields are defined in terms of pairs of points rather than single points. We will do that as follows: 

DEFINITION: Let $U$ be some subset of $M$. If $\prec$ is some causal relation on $M$, then $G (U, \prec)$ is a set of all causal relations $\prec^*$ such that for all $p \in U$ , $p \prec q \Leftrightarrow p \prec^* q$  and $q \prec p \Leftrightarrow q \prec^* p$ for any point $q$ on a manifold, regardless whether $q$ is an element of $U$ or not. 

DEFINITION: Let $U$ be some subset of $M$. If $a \colon S \times S \rightarrow \mathbb{R}$ is some real valued function on the set of pairs of points of $S$, then $A (U, \prec)$ is a set of all other functions $b \colon S \times S \rightarrow \mathbb{R}$ such that for all $p \in U$ , $a(p,q)=b(p,q)$  and $a(q,p)=b(q,p)$ for any point $q$ on a manifold, regardless whether $q$ is an element of $U$ or not. 

THEOREM: If $\prec_1 \in G (U, \prec_2)$ then $\prec_2 \in G (U, \prec_1)$

PROOF: If $\prec_1 \in G (U, \prec_2)$ then for any $p \in U$, $p \prec_1 q \Leftrightarrow p \prec_2 q$. This is equivalent to $p \prec_2 q \Leftrightarrow p \prec_1 q$ which by definition means $\prec_2 \in G (U, \prec_1)$.   

THEOREM: If $\prec_2 \in G (U, \prec_1) $ then $G (U, \prec_1) = G(U, \prec_2)$

PROOF: Suppose $\prec_3 \in G (U, \prec_2)$. Then for any $p \in U$, $p \prec_3 q \Leftrightarrow p \prec_2 q$ for all $q$. But since $\prec_2 \in G(U, \prec_1)$ we also know that for any $p \in U$, $p \prec_2 q \Leftrightarrow p \prec_1 q$ for all $q$. Thus, putting these two together we get $p \prec_3 q \Leftrightarrow p \prec_1 q$ for all q. Thus we have shown that $\prec_3 \in G (U, \prec_2) \Rightarrow \prec_3 \in G (U, \prec_1)$ which means $G(U, \prec_2) \subset G (U, \prec_1)$ whenever $\prec_2 \in G (U, \prec_1)$. But from the previous theorem we know that since $\prec_2 \in G (U, \prec_1)$ we have $\prec_1 \in G (U, \prec_2)$. Thus, we can rearrange $\prec_1$ and $\prec_2$ in the previous statement and say $G(U, \prec_1) \subset G (U, \prec_2)$. Putting these results together, we have $G (U, \prec_1) = G (U, \prec_2)$

Thus, a causal set version of {\cal A} will be

For bosonic case, we have 

\bea {\cal A} (\prec, \phi_1, . . . , \phi_n , a_1 , . . . , a_m ; T_1, T_2) = \nonumber \eea
\bea = \Big(\sum_{ \{ \prec^*, a_1^* , . . ., a_m^*, \phi_1, . . . , \phi_n \vert a_k^* \in A (T_1 , \prec) \; and \; \prec^* \in G(T_1 , \prec) \} } e^{i S (\prec, \phi_1, . . . , \phi_n, a_1, . . . , a_m)}\Big)^{-1} \times \nonumber \eea
\bea \times \Big(\sum_{ \{ \prec^*, a_1^* , . . ., a_m^*, \phi_1, . . . , \phi_n \vert a_k^* \in A (T_1 \cup T_2, \prec) \; and \; \prec^* \in G(T_1 \cup T_2 , \prec \} }  e^{i S (\prec, \phi_1, . . . , \phi_n, a_1, . . . , a_m)}\Big) \eea

Now if we include fermions we will get

\bea {\cal A} (\prec, \phi_1, . . . , \phi_n , a_1 , . . . , a_m, ; T_1, T_2) = \nonumber \eea
\bea = \Big(\sum_{ \{ \prec^*, a_1^* , . . ., a_m^*, \phi_1, . . . , \phi_n \vert a_k^* \in A (T_1, \prec) \; and \; \prec^* \in G(T_1, \prec \} } \lambda (\prec, a_1, . . . , a_m, \phi_1, . . . , \phi_n) \times \nonumber \eea
\bea \times  \big(\bigwedge \hat{\psi_s} \big) \cdot e^{i S (\prec, \phi_1, . . . , \phi_n, a_1, . . . , a_m)}\Big)^{-1} \times \nonumber \eea
\bea \Big(\sum_{ \{ \prec^*, a_1^* , . . ., a_m^*, \phi_1, . . . , \phi_n \vert a_k^* \in A (T_1 \cup T_2, \prec) \; and \; \prec^* \in G(T_1 \cup T_2, \prec \} } \lambda (\prec, a_1, . . . , a_m, \phi_1, . . . , \phi_n) \times \nonumber \eea
\bea \times \big(\bigwedge \hat{\psi_s} \big) \cdot e^{i S (\prec, \phi_1, . . . , \phi_n, a_1, . . . , a_m)}\Big)\eea  

Here it is understood that while formally $\lambda$ depends on all fields, it actually only depends on the ones that are part of the definition of fermion. It is simply that since both fermionic and bosonic fields are now defined in terms of functions on both $S$ and $S \times S$ I formally written that $\lambda$ depends on all of them in order to save space.

We can likewise define a "discritized shaddow" on a causal set. Of course, causal set is discrete to start with. So we will replace the word "discritized" with the word "coarse grained", where "coarse graining" is a well known operation on causal set that involves selection of some subset of points that is "dense" enough to approximate any other point in a causal set. 

DEFINITION: Let $T$ be a subset of a causal set $S$.  Its past $\tau$-subset is a union of its past $\sigma$-shaddows corresponding to all $\sigma \leq \tau$ 

DEFINITION: Let $T$, $U$, and $V$ be some subsets of a causal set $S$, and suppose $U \subset V$. We say that $U$ is coarse-grained past $\tau$-shadow of $T$ and $V$ is coarse-grained past $\tau$-lattice based on $T$ if  the following statements are true: 

a)Every element of $V$ is causally after at least one element of the $\tau$ - past shaddow of $T$ and causally before at least one element of $T$

b)Every single element of past $\tau$-shaddow of $T$ is to the past of at least one element of $V$. Likewise, every single element of $T$ is to the future of at least one element of $V$

c)If the number of points contained in past $\tau$-submanifold of $T$ is $I$, then the number of elements of $V$ is less than $I/v_1$ and greater than $I/v_2$

d)If $p$ and $q$ are two elements of $V$ and the number of elements of a causal set that are to the future of $p$ and to the past of $q$ is greater than $n_1$ (this places lower bound on volume of space bounded by the two lightcones and hence the Lorentzian distance) then there are at least $n_2$ points $r$ that are to the future of $p$ and to the past of $q$ (where my definition of past and future includes the fact that they are timelike-separated from $p$ and $q$ )

e)If $r \in U$ then there exist $s$ that is to the causal past of $r$ such that $r$ and $s$ are connected by at least one future directed chain of points that has less than $m_1$ points (i.e. an upper bound on Lorentzian distance) but at the same time we have MORE than $m_2$ points that are part of the shaddow of $T$ and are both to the causal future of $s$ and to the causal past of $r$. This criteria assures that the reference frame defined by discritized shaddow approximates the one defined by actual shaddow and rules out the issues of lightcone singularities.

We will likewise introduce a definition of "future" discritized shaddow:

DEFINITION: Let $T$ be a subset of a causal set $S$.  Its future $\tau$-subset is a union of its future $\sigma$-shaddows corresponding to all $\sigma \leq \tau$ 

DEFINITION: Let $T$, $U$, and $V$ be some subsets of a causal set $S$, and suppose $U \subset V$. We say that $U$ is coarse-grained future $\tau$-shadow of $T$ and $V$ is coarse-grained future $\tau$-lattice based on $T$ if  the following statements are true: 

a)Every element of $V$ is causally after at least one element of the $\tau$ - future shaddow of $T$ and causally before at least one element of $T$

b)Every single element of future $\tau$-shaddow of $T$ is to the future of at least one element of $V$. Likewise, every single element of $T$ is to the past of at least one element of $V$

c)If the number of points contained in future $\tau$-submanifold of $T$ is $I$, then the number of elements of $V$ is less than $I/v_1$ and greater than $I/v_2$

d)If $p$ and $q$ are two elements of $V$ and the number of elements of a causal set that are to the past of $p$ and to the future of $q$ is greater than $n_1$ (this places lower bound on volume of space bounded by the two lightcones and hence the Lorentzian distance) then there are at least $n_2$ points $r$ that are to the past of $p$ and to the future of $q$ (where my definition of past and future includes the fact that they are timelike-separated from $p$ and $q$ )

e)If $r \in U$ then there exist $s$ that is to the causal future of $r$ such that $r$ and $s$ are connected by at least one future directed chain of points that has less than $m_1$ points (i.e. an upper bound on Lorentzian distance) but at the same time we have MORE than $m_2$ points that are part of the shaddow of $T$ and are both to the causal past of $s$ and to the causal future of $r$. This criteria assures that the reference frame defined by discritized shaddow approximates the one defined by actual shaddow and rules out the issues of lightcone singularities.

Now we do the same tricks to introduce versions a, b and c of constraint 1 for causal sets, which we will denote as 1* to distinguish it from manifold case: 

CONSTRAINT 1* VERSION A: Let $S$ be causal set. If we define measure $\mu$ to be integer-based, defined in terms of simple count of relevent multiplets of points, then the following statement is true: For every number $n$, if we have points $q_1, . . . , q_n$,

\bea \forall \tau> \Delta \big[p_0- \epsilon_1< \mu \{q_1, . . . , q_n \vert p_0-\epsilon_2< p({\cal F}; S_{past}(\{ q_1 , . . . , q_n \}, \prec, \tau), \{ q_1, . . . , q_n \} ) < p_0 + \epsilon_2 \} < \nonumber \eea
\bea < p_0 + \epsilon_1 \big]  \eea

CONSTRAINT 1* VERSION B: Let $S$ be causal set. If we define measure $\mu$ to be integer-based, defined in terms of simple count of relevent multiplets of points, then the following statement is true: For every number $n$, if we have points $q_1, . . . , q_n$,

\bea \forall \tau > \Delta \big[p_0- \epsilon_1< \mu \{q_1, . . . , q_n \vert \; if \; T \; is \; \delta \; shift \; of \; S_{past}(\{ q_1 , . . . , q_n \}, \prec, \tau) \; \nonumber \eea
\bea then \; p_0-\epsilon_2< p({\cal F}; T, \{ q_1, . . . , q_n \} ) < p_0 + \epsilon_2 \} < p_0 + \epsilon_1 \big]  \eea

CONSTRAINT 1* VERSION C: Let $S$ be causal set. If we define measure $\mu$ to be integer-based, defined in terms of simple count of relevent multiplets of points, then the following statement is true: For every number $n$, if we have points $q_1, . . . , q_n$,

\bea \forall \tau > \Delta \big[p_0- \epsilon_1< \mu \{q_1, . . . , q_n \vert \; if \; T \; is \; \delta \; coarse-graining \; of \; S_{past}(\{ q_1 , . . . , q_n \}, \prec, \tau) \; \nonumber \eea
\bea then \; p_0-\epsilon_2< p({\cal F}; T, \{ q_1, . . . , q_n \} ) < p_0 + \epsilon_2 \} < p_0 + \epsilon_1 \big]  \eea

Finally, constraint 2, which guarantees a local continuity of fields, translates as follows:

CONSTRAINT 2* FOR SCALAR FIELDS: Let S be a causal set and let $\phi \colon S \rightarrow \mathbb{C}$ be scalar field. For every $p \in S$ there exist $q \prec p$ such that the following is true:

a)$ \sharp \{ r \vert q \prec r \prec p \} \geq \rho$ where $\sharp$ stands for number of elements

b)For any $r$ satisfying $q \prec r \prec p$, $\vert \phi (r) - \phi (p) \vert < \sigma$

ONSTRAINT 2* FOR CAUSAL SET VERION OF VECTOR FIELDS: Let S be a causal set and let $a \colon S \times S \rightarrow \mathbb{R}$ be a function interpretted as vector field. Then for every $p \in S$ there exist $q \prec p$ such that the following is true:

a)$ \sharp \{ r \vert q \prec r \prec p \} \geq \rho$ where $\sharp$ stands for number of elements

b)For any $r_1$, $r_2$ and $r_3$ satisfying $q \prec r_k \prec p$, $\vert (a(r_1 , r_2) + a (r_2 , r_3) + a (r_3, r_1) \vert < \sigma$ 

The rest is just a carbon copy of what we did for bosonic case: 

The probability density will be  

\bea p (\prec, \phi_1, . . . , \phi_n , a_1 , . . . , a_m ; T_1, T_2 ) = \vert {\cal A} (\prec, \phi_1, . . . , \phi_n , a_1 , . . . , a_m; T_1, T_2 ) \vert ^2 \eea

And, based on this probability, I impose either version A (hard version) or versions B or C (easy versions) of constraint 1* on the collection of all possible scenarios where $T_2$ is a shaddow of $T_1$: 
 
Finally, the causal set version of localizing fermions into a world path is the following: 

CONSTRAINT 3*: We still have fields $\chi_p$, $\chi_a$ and $u_a^{\mu}$ throughout space time. But the set in which $\chi_p$ and $\chi_a$ is non-zero looks like a set of sequences of points that oscillate between future and backward direction: $r_{1} \prec . . . \prec r_{n_1} \succ . . . \succ r_{n_1+ n_2} \prec . . . \prec r_{n_1 + n+2 +n_3} \succ . . . $ such that one can NOT find point $s$ and number $k$ for which either $r_k \prec s \prec r_{k+1}$ or $r_k \succ s \succ r_{k+1}$ (in other words, we don't "skip" over any points, making the sequence of points curve-like) and total number of points in each such sequence is limitted by $M$ (by making $N$ greater than the separation of earliest and latest points of causal set, but not too much greater, I will assure that most of each sequence is taken up by future directed pieces making the picture matter-dominated). Total number of these sequences, on the other hand, is bounded by $N$ (this assures that they don't "fill" an entire causal set). If I will orient them in a particular way, these curves have the following property: each of these curves starts at a point that is not to the future of any other point, and ends at a point that is not to the past of any other point (in other words, it starts at a "past" surface of space-time and ends at a "future" point of space-time) The total length of each curve is bounded by $M$ (by making $M$ greater than the separation of past and future surfaces of the universe, but not too much greater, I will assure that most of each curve is taken up by future directed pieces). On the future directed piece of each curve, $\chi_a=0$ while on the past directed piece of each curve $\chi_p=0$. The future directed pieces are interpreted as particles, while the past directed pieces are interpreted as antiparticles. Their junctions are interpreted as pair creation and pair annihilation. 

One important aspect of the above is that one of the "fields" I am including is a causal relation $\prec$. This can be done while still retaining the background topology due to the following two facts:

1)Floating causal relation $\prec^*$ co-exists with fixed causal relation $\prec$. The former is a field while the latter provides a topological background. 

2)The range of fluctuations of $\prec^*$ is limitted by a constraint that $\prec^*$ agrees with $\prec$ as long as one of the two points is either an element of $\{ q_1 , . . . , q_n \}$ or its shaddow.

On the one hand, for the two reasons described above we do get well defined topological background as well as well defined quantum field theory of gravity on that background. On the other hand, however, it is possible that the actual numbers we will get will be too chaotic. The reason for this is that since we let the defining parameter of a shaddow, $\tau$ to be very large, we still have fluctuations of gravitational field on a macroscopic level.  In case of non-gravitational fields at least we have some assurance that is discussed in section 2. But in case of gravity we lose confidence since we no longer assume fixed background that we have assumed back then. For that reason I will provide some further constraints of the theory that might be a "safety net". Since it is possible that after the numerical work the theory will work without that safety net, for now I will view these extra constraints as optional.

The additional constraint that I propose will be similar to the "tunnel" discussed in chapters 5 and 6 of [13]: I will impose a restriction that gravitational field defined by $\prec^*$ approximates gravitational field defined by $\prec$. In other words, I would like to define a "neighborhood" of $\prec$, which I will call $n (\prec, \epsilon, N)$  (where N is the a low bound on a distance scale imposed in order to avoid unwanted discrete effects) and in the above definitions I will replace $G(T, \prec)$ with $H(T, \prec, \epsilon, N)$ where $H(T, \prec, \epsilon, N) = G(T, \prec) \cap n(\prec, \epsilon, N)$ 

In light of the fact that classical Einstein equation can be derived by means of variation of an action with respect to $g_{\mu \nu}$, ideally I would like to define $n(\prec)$ in terms of small variation of the same. However, imposing separate restrictions for each of the choice of $\mu$ and $\nu$ is not a relativistically covariant procedure, which means it is not well defined for causal set. However, due to the fact that length of geodesic depends linearly on $g_{\mu \nu}$ this becomes an easy replacement. Timelike geodesic is the only one I have definition for, so that is the one I will use. We have to take into account the fact that since $\prec$ and $\prec^*$ are distinct, some pairs of points are related by one causal relation and not by the other, which means that only one of the two geodesic segments whose length we want to compare is present. The way to deal with this case is to notice that if $g_{\mu \nu} \approx g^*_{\mu \nu}$ then whenever Lorentzian distance according to these two metrics differ in sign, they should both be close to $0$. Thus, a constraint to impose is that if $p \prec q$ AND their distance is large according to $\prec$, then $p \prec^* q$. On the other hand, we still have to impose the constraint regarding comparable geodesic length. This, too, requires the scenario where two points have large enough Lorentzian distance according to relevent partial orders, since we need continuum-looking picture in order to talk about ratios in a meaningful way. Thus, we can combine these two constraints into one constraint and simply say that if $p \prec q$ and they are far enough from each other according to $\prec$, then $p \prec^* q$ and their distance according to $\prec^*$ will be comparable to the one according to $\prec$ :

DEFINITION: Let $\prec$ be a partial order on a causal set $S$. Let $\epsilon$ be some small real number and let $N$ be an integer. Let $\prec^*$ be some other partial ordering on $S$. Then $\prec^*$ is an element of $n_{\tau}( \prec, \epsilon, N)$ if for any points $p \prec r_1 \prec . . . \prec r_N \prec q$ we can find points $s_1, . . . , s_M$ satisfying $p \prec s_1 \prec . . . \prec s_M \prec q$ where $M \geq (1- \epsilon) N$. Likewise, for any points $p \prec^* r_1 \prec^* . . . \prec^* r_N \prec q$ we can find points $s_1, . . . , s_M$ satisfying $p \prec s_1 \prec . . . \prec s_M \prec q$ where $M \geq (1- \epsilon) N$

The reason we have put $n_{\tau}$ rather than $n$ is that this was a way of defining $n$ in terms of Lorentzian distances, $\tau$. If we want, we can instead use a definition in terms of volumes. If we have two causally related points $p$ and $q$, we can single out a set of points $r$ satisfying $p \prec r \prec q$. Geometrically, this looks like a space bounded by future part of a light cone of $p$ and past part of light cone of $q$. This space is called Alexandrov set. The volume of that space is proportional to the number of points contained in it. Since in locally flat region it is proportional to the Lorentzian distance taken to the fourth power, we can rewrite the above definition replacing Lorentzian distance with volumes: 

DEFINITION: Let $\prec$ be a partial order on a causal set $S$. Let $\epsilon$ be some small real number and let $N$ be an integer. Let $\prec^*$ be some other partial ordering on $S$. Then $\prec^*$ is an element of $n_{V}( \prec, \epsilon, N)$ if whenever there are more than $N$ choices of $r$ satisfying $p \prec r \prec q$, we also have more than $(1-\frac{\epsilon}{4})N$ choices of $s$ satisfying $p \prec^* s \prec q$. Likewise, if we have more than $N$ choices of $r$ satisfying $p \prec^* r \prec q$, we also have more than $(1-\frac{\epsilon}{4})N$ choices of $s$ satisfying $p \prec s \prec q$.  

The reason we used $\frac{\epsilon}{4}$ in the above definition instead of $\epsilon$ is that I would like to make sure that geodesic length, as opposed to the volume, deviates by order of $\epsilon$ since the former rather than the latter linearly depends on the metric. I didn't bother doing similar conversion for $N$ since the latter is just intended to put some low bound on the number of points to avoid discreteness effects, which means that it doesn't make a physical difference whether it refers to lengths or volumes.  

Of course, this still doesn't adress the renormalization issue. But since causal set theory can be done numerically, we can still claim that physics is well defined. 

Before we finish a section on causal set theory it is important to mention one problem of causal set model that might possibly be answered by the approach to quantum mechanics discussed in this paper. The problem I would like to discuss is this: if our space time is a causal set, how can we explain its manifoldlike structure? The statistical properties of causal sets were widely investigated. One of the more common models is "dynamical percolation" (see [5]). According to this model, we dynamically add new points, and each time we add a point we make a random choice: for each of the existing points, a new point will be "after" an existing point with probability $p$, and unrelated to the existing point with probability $1-p$. It had been shown that if we perform this process, then with 100 percent probability we will get an infinite sequence of "posts". By "post" we mean a causal set version of big bang: a point that is causally related to every single other point in a causal set. Unless $p$ is very small, these posts (or big bangs) will be separated only few points from each other, which is clearly not our version of space time. While the dynamical percolation model is not the only model of generating of causal sets that has been investigated, it is safe to say that so far none of the models have been proposed that would dynamically reproduce a manifoldlike structure.

I propose that our approach might be a candidate for solving the above puzzle for two reasons:

1)According to our approach, all dynamics is being replaced with space-time geometry. I obtain the desired physics by "precluding" unwanted four dimensional histories. This means that I am no longer obligated to come up with a dynamical laws that generate manifoldlike structure. I can simply preclude the ones that are not manifoldlike, as long as I make an adequate four dimensional definition of manifoldlike. 

While the above can in principle work by itself, it looks too much like cheating. For this reason, I will attempt to use the above in conjunction with the following:

2)Motivated by a common understanding that gravity is both geometry and a field, I will claim that manifoldlike geometry is a result of specific physics rather than its stage. In particular, I will make Lagrangian-preclusion version of a statement that fermionic field is a physical force that "twists" gravity into manifoldlike structure. 

Thus, in my future research I will first start from number 2 and hope that enforcing 2 by using preclusion principle will somehow provide a manifoldlike structure for me; only after that, if that clearly won't work, I might apply part 1 directly to manifoldlike structure if I become desperate. 

The way I propose to use fermionic field is as follows: as we have said before, I have four vector fields as part of my definition of fermionic field. In the context of a causal set, these four vector fields are represented in terms of four functions of the form $S \times S \rightarrow \mathbb{R}$ : $a_0(p,q)$, $a_1(p,q)$, $a_2(p,q)$ and $a_3(p,q)$. We can define Lagrangian density at a point $p$ as follows:

DEFINITION: Let $p$ be an element of causal set. A Lagrangian density at a point $p$ is ${\cal L} = f(n(\delta))$ where f is some real valued function while $n(\delta)$ is a number of choices of points $q$ and $r$, satisfying $\vert a_k(p,q) \vert < \delta$ and $\vert a_k (p,r) \vert < \delta$, for which exactly ONE of the following two statements is true:

a)$q \prec r$ or $r \prec q$

b)$(a_0 (p, q) - a_0 (p,r))^2>(a_1 (p, q) - a_1 (p,r))^2+(a_2 (p, q) - a_2 (p,r))^2+(a_3 (p, q) - a_3 (p,r))^2 $

Normally, one would expect a and b to go together. Thus, by saying that exactly ONE of these is true we are basically saying that a choice of points is affected by curvature: either coordinate-wise they look like causally related while they trully aren't, or the other way around. A condition that $a_k(p,q) < \delta$ and $a_k (p,r) < \delta$ basically means we single out a coordinate neighborhood around $p$ that looks like a square. The neighborhood is defined in terms of coordinates as opposed to Lorentzian distance in order to avoid singularities near light cone. So $n(\delta)$ counts the number of pairs of points that are affected by curvature in $\delta$-neighborhood of $p$. Now if I define $f$ in such a way that $f(0)=f'(0)=0$ while it rapidly increases away from $0$, then principle of least action would pick out a scenario where neighborhoods of all points are flat, which is a rough version of definition of manifold. Since we allow quantum fluctuations, this would fluctuate away from the manifold, but we can still reasonably expect that if we make $f$ increase fast enough we will get some approximation to the manifold.

If we didn't define preclusion principle the way we did, one of the possible obstacles could have been the fact that we had to define measure in a very clever way in order to define Grassmann integral. This is another side of a coin of the following issue: our intention was to define path integral to be an equivalent of the Grassmann integral that is already known. Since Grassmann integral, as we know it, has no meaning outside of an integral, thus, while we are free to imagine some picture that has some geometrical meaning, it won't give us any new physical information. However, this situation has changed when we have invented a view according to which every single thing that we take path integral over, has to be physically realizeable. According to this principle, when we made our new definition of Grassmann numbers, we accomplished two different tasks:

1)We came up with better definition of integral

2)We claim that these geometrical objects that we defined have to be PHYSICALLY realized in one of the parallel worlds

While part 1, as I said, has no physical implications other than a mathematical interest, part 2 does give us a physical inside. Furthermore, the cleverly adjusted probability distribution only plays a role in part 1. On the other hand, as far as part 2 is concerned, we can simply discritize the values of two-point functions by some very small interval and claim that every single assignment of these discrete values to pairs of points is physically realized, except for the precluded ones. This basically means that we have a new definition of probability, which is no longer "weird". While the "weird" definition of probability is used for integral, the non-weird one is used for actually realizing these possible states into physical world. So, due to the non-weird definition of probability, fermionic fields helped us provide a manifold structure.

Thus, in my future research I will first start from number 2 and hope that enforcing 2 by using preclusion principle will somehow provide a manifoldlike structure for me; only after that, if that clearly won't work, I might apply part 1 directly to manifoldlike structure if I become desperate.

\bigskip

\noindent{\bf 7. Conclusion}

After having read this paper, a natural question arises that may be there was some kind of cheating given that a paper adresses questions from interpretation of quantum mechanics to defending both of the opposing views on quantum gravity to some issues like matter dominated universe (end of section 4) on a side. Indeed, a cheating did occur. By postulating a "preclusion principle" one can essentially preclude anything one likes. For example, one can explain why the trees grow green leafs by "precluding" every single scenario where leafs have non-green color with an exception of a possibility of them first being green and then changing color as well as possibility of them being attached to the tree by hand. This basically means that in this paper we haven't done anything other than re-stating already known observations in a clever way and calling it a theory, which is precisely why so many things are explained so easilly. However, to the defense of what was done one might quote classical general relativity. In that theory, due to the fact that dynamics is being replaced by geometry, one can also argue that basically a list of "allowed" dynamics is being listed, without any apparent mechanism behind them. So the question arises: where can one draw a line between classical general relativity and preclusion theory of leafs being green, and on what side of a line does this paper fall? 

One answer that we can come up with is this: we can quote an idea expressed by Mach and others that a good theory is the one that is simplest possible one. For the purposes of this paper, we can support Mach's principle by using the works of another philosopher, Hume, who argued that there might not be such a thing as a cause, but only our observation of patterns. When we say that "A happens because of B" we can translate it into passive language and say that "A always happens after B, so since in this particular case B have happened we know that A will happen as well". Here we didn't really explain why "A always happens after B"; rather, it was just an observation. But in this case, why couldn't have we simply observed A happening on Monday and then say it happened on Monday because that is part of the observation? The answer is that "A always happens after B" is simpler than a list of days when A happened, and simpler theory is automatically a truer one. Now lets go back to the example of photosynthesis. Even if we use a preclusion principle as our explanation, we still have to acknowledge the existence of photosynthesis, simply because the phenomenon is "there"; the only thing that we deny is its key role of explaining why the leaf is green. Furthermore, if we acknowledge that photosynthesis is there, we have to acknowledge the obvious colorary, namely that the leafs would of been green even if non-green colors were not precluded. Thus, we have two candidates for a theory: one is photosynthesis alone while the other is photosynthesis+preclusion. In this case, photosynthesis-alone theory, being a simpler theory, wins. On the other hand, in case of classical gravity, a theory of differential equations proves that a given set of equations can not be reduced to a smaller set of equations, and that is precisely what defends general relativity against this kind of criticism. 

This criteria, however, has its own problem. According to this criteria, in order to defend any theory against the green leaf analogy one has to simply show that there is no simpler theory. This means that if one takes any un-answered question, one can simply take advantage of the fact that it is un-answered to argue that preclusion principle that would "force" an answer to this question is a best possible theory. One can then try a different answer: Apart from the principle that "simpler theory is a better theory" we can argue something else: can a preclusion principle be modeled in terms of parallel universes? In case of quantum phenomena the question is yes; we can imagine all possible parallel universes where every single non-precluded scenario occurs. On the other hand, in case of biology the answer is no, since we know that cells and even DNA molecules are classical objects, hence behave in only one way no matter how many universes we consider. 

While this answer can be satisfactory to some, it wouldn't be satisfactory to others. After all, while it no longer asks us to forget about "all" of the unsolved problems, it "only" asks us to forget about the ones pertaining to interpretation of quantum mechanics, in other words, the most interesting ones. Thus, people who are not interested in interpretation of quantum mechanics and are more phenomenologically oriented, can use my very argument not to worry about it. On the other hand, the few people who are interested in interpretation of quantum mechanics have to first come up with their own version of stricter criteria of a good theory, so that, despite the above argument that the theory without interpretation is "good" by a looser criteria, it would no longer be good by a stricter criteria. And then, of course, a successful project in interpretation of quantum mechanics would come up with a theory that would meet that stricter criteria. Since a criteria is an axiom rather than a theorem, actual choice of stricter criteria is a philosophy rather than hard science. For that reason, I personally am in favor of trying to experiment with a lot of different philosophies, and for each find their own theory in order to "satisfy every view". 

One of the long list of the philosophies should, of course, be a philosophy of a typical physicist, that a value of a theory is its ability to make predictions. This might, in fact, be an alternative to Mach's principle. A "simple theory" that "A always happens after B" is capable of making predictions, while a "complicated theory" which lists all the days when A have happened in the past, is not. Thus, an avenue to explore is to do some quantitative computations by using the theory presented in this paper and explore what kind of predictions it can make. While none of this have been done, one qualitative guess we can make is this: if it turns out that "easier" version rather than "harder" version of constraint 1 should be used, the consequence of that is that different field fluctuations would have a tendency to pick out common frame. Of course, there are other reasons why they do pick common frame, namely all the different processes that lead to formation of planets and other macroscopic objects after the big bang. But the model presented in this paper, by showing an additional reason for the same thing, might potentially imply that the formation processes should occur faster. Again, this is something I can't know until I perform calculations. But in case that calculations will confirm the qualitative argument this would provide one avenue of testing the theory.

There is also a different problem with this paper that follows the same lines: how do I know that a theory is relativistically covariant if anything and everything can be turned into relativistically invariant? For example, suppose Klein Gordon Lagrangian was non-relativistic, ${\cal L} = (\partial_0 \phi)^2$. Then we can simply rewrite it by saying that there exist some timelike vector field $V$ for which ${\cal L} = (V^{\mu} \partial_{\mu} \phi)^2$. In our work we have done similar trick: in our discussion of "easier" version of constraint 1, we have relied on the fact that only in some frames the fields don't fluctuate too much. We further notice that it doesn't have to be "some frames". It is possible that in all frames fields fluctuate by huge amounts; and then we used constraint 2 to rule that out. So does that mean that we basically postulated ether in constraints 1B and 2? On the one hand, I introduced constraint 2 in relativistically invariant way since when I said "there exist a certain choice of a neighborhood" the statement "there exist" is either true in all frames or false in all frames. At the same time, however, whatever I say "exist" can be construed to correspond to reference frame, i.e. reference frame of boundaries of a certain region. Thus, the fact that "there exists" were used as opposed to "for every" implies violation of relativity. This is similar to the example of Klein Gordon equation just described: on the one hand, $V^{\mu} \partial_{\mu} \phi$ is covariant, but on the other hand, the fact that $V^{\mu}$ was specified implies violation of relativity. 

The ultimate question raised by these examples is how do we distinguish a physical system from ether? Any physical system can set a preferred frame. For example if we have a simple lab experiment with few charges and magnets, then if we hide them we can deceive a student into thinking that translational and boost symmetries are violated. So the way that these fields, determined be lab equipment, are different from the ether is that they are changeable, while ether is not. But then what if we have something in a gray area: a very heavy Higgs field that is changeable, but very hard to change? On the other hand, what about the fact that due to the fact that gravity is quantized, a geometry is no longer regarded as unchangeable. Since according to causal set approach I identify gravity with a causal structure, this means that lightcone causality is not unchangeable, which contradicts one of the key axioms Einstein proposed for relativity. 

The answer is the same as the answer for the previously raised question regarding good theory. Namely, un-like what most physicists will tell you, the criteria whether something is relativistically invariant or not is a question of philosophy rather than hard science. Of course, relativity have been observed in the lab, and such observations have to be predicted by any proposed theory. However, a lot of discrete theories might violate relativity by postulating a specific microscopic structure, and then explain why it is observed in the lab despite being violated on microscopic level. Despite the fact that these theories don't predict deviations from relativity in the lab, people still don't like them. This means that belief in relativity reached a level of aesthetics, and went well beyond the observational criteria. And that aesthetic criteria is where we are no longer sure what the rigurous definition of "relativity" is.  

%\newpage

\end{document}